\documentclass[a4paper,12pt]{article}

\newcommand{\prop}{\widetilde{\Delta}_A}
\newcommand{\fakt}{f_A}
\newcommand{\RR}{{\rm R}}
\newcommand{\RRR}{\RR^{1,3}}
\newcommand{\RRRR}{\RR^{4}}
\newcommand{\CC}{{\cal C}}
\newcommand{\SEP}{{\cal C}'}
\newcommand{\SMA}{S_A}
\newcommand{\GFA}{G_A^{(n)}}
\newcommand{\GFAT}{G_A^{(2)}}
\newcommand{\SE}{\Pi_A^*}

\newcommand{\Iq}{\tilde{I}}
\newcommand{\zq}{\tilde{z}}
\newcommand{\qc}{\tilde{q}}
\newcommand{\zqr}{\tilde{x}}
\newcommand{\zqi}{\tilde{y}}

\title{Framework for finite alternative theories to a quantum field theory. II-Unitarity}
\author{Marijan Ribari\v c and Luka \v Su\v ster\v si\v c\thanks{Corresponding author. Phone +386 1 477 3258; fax +386 1 423 1569; electronic address: \tt luka.sustersic@ijs.si\rm} \\Jo\v zef Stefan Institute, p.p.3000, 1001 Ljubljana, Slovenia}

\date{}

\begin{document}

\maketitle
\begin{abstract}
We generalized the 't Hooft-Veltman method of unitary regulators to put forward a path-integral framework for finite, alternative theories to a given quantum field theory. And we demonstrated that the proposed framework is feasible by providing a finite alternative to the quantum field theory of a single, self-interacting real scalar field. Here we give two properties of self-energy that make the corresponding scattering matrix unitary. We show that the perturbative self-energy has these two properties at least up to the second order in the coupling constant.
\end{abstract}

\bigskip
Suggested short title: Framework for finite alternative theories ...

\bigskip
\bigskip

\section{Introduction}

In part I of this paper \cite{mi001}, using the \it path-integral formalism, \rm we considered a new, covariant, Lagrangian-based framework for constructing \it finite, alternative theories to a given quantum field theory \rm (QFT). Such theories provide solutions to the problem of QFT ultraviolet divergencies by effecting a non-perturbative regularization without resorting to discrete space-time, additional space-time dimensions, formal auxiliary parameters, or auxiliary particles with negative metric or wrong statistics. So such a regularization is realistic in the sense of Pauli and Villars \cite{Pauli}.

In this paper we consider for a scalar case the problem of perturbative unitarity within the proposed framework. To this end we use the cutting equations approach (Sec.~9.6 in Ref.~\cite{Sterman}) that was put forward by Veltman (\cite{Velt1}, and Ch.~8 in Ref.~\cite{Velt2}) and t'Hooft and Veltman (Sects.~4-7 in Ref.~\cite{Hooft}).

In Sec.~\ref{alterpropag}, we introduce the considered scalar case.

In Sec.~\ref{partcont}, we point out the properties of an alternative self-energy that make the corresponding alternative scattering matrix model quantum scattering of scalar particles of the same mass.

In Sec.~\ref{perturunitar} we give the properties of an alternative Lagrangian that imply the perturbative unitarity of the alternative scattering matrix by the cutting equations approach.

In Sec.~\ref{pertprop}, we show explicitly up to the second order of the coupling constant that a particular alternative Lagrangian has such properties.

In Sec.~\ref{komentar}, we comment on proving perturbative unitarity of the particular alternative scattering matrix also up to all orders.

\section{A scalar case}

\subsection{Properties of an alternative propagator}
\label{alterpropag}

As an example let us stay with the alternative theory to the QFT of a single real scalar field with $\phi^4$ interaction, and the corresponding alternative scattering matrix $\SMA$ we considered in Ref.~\cite{mi001}. The corresponding perturbative $\SMA$-matrix is defined in terms of perturbative expansions of the alternative Green functions $\GFA$ by the Lehmann-Symanzik-Zimmermann (LSZ) reduction formula. To obtain perturbative expansions of the alternative momentum-space Green functions $\GFA$, we can use, e.g.~the following recipe \cite{mi001}: (I)~We write the interaction Lagrangian as
\begin{equation}
   - {\lambda\over 4!} \phi^4 - {Z^2 \lambda_0 - \lambda \over 4!} \phi^4 - {Z-1\over 2} (\partial_\mu \phi)^2 
           - {Z m_0^2 - m^2\over 2} \phi^2 \,, 
   \label{interLagr}
\end{equation}
where $\lambda$, $Z$, $\lambda_0$, $m$, and $m_0$ are real coefficients that depend on $\lambda$ so that $Z = 1$, $\lambda_0 = 0$, and $m_0 = m$ for $\lambda = 0$; and we use the $(-1, 1,1,1)$ metric. (II)~In perturbative expansions of the QFT, Minkowskian momentum-space Green functions corresponding to the interaction Lagrangian (\ref{interLagr}), we replace the spin 0 Feynman propagator with an alternative propagator
\begin{equation}
   \prop(k^2) = (k^2 + m^2 - i\epsilon )^{-1} \fakt(k^2 - i\epsilon) \,, \qquad \epsilon \searrow 0 \,, \quad k \in \RRR \,,
   \label{propagator}
\end{equation}
where: (a)~the alternative regularizing factor $\fakt(z)$ is an analytic function of $z \in \CC$ except somewhere along the segment $z \le z_d < -9m^2$ of the real axis; (b)~$\fakt(-m^2) = 1$; (c)~$\fakt(z)$ is real for all $z > z_d$, so $\fakt(z^*) = \fakt^*(z)$; (d)~we can estimate that for all $z \in \CC$, 
\begin{equation}
   | \fakt(z) | \le a_0 ( 1 + | z |^{3/2} )^{-1} \,, \qquad a_0 > 0 \,;
   \label{faktest}
\end{equation}
and that for any real $z_0 > z_d $ the derivatives $\fakt^{(n)}(z)$ of $\fakt(z)$ are such that
\begin{equation}
   \sup_{ z \in \CC, \Re z \ge z_0} (1 + |z|^{3/2}) (1 + |z|^n) | \fakt^{(n)}(z) | < \infty \,, \quad n = 1, 2, \ldots ;
   \label{ocenaodvoda}
\end{equation}
and (e)~we can make the coefficients of $\fakt(z)$ depend on a positive cut-off parameter $\Lambda$ as specified in Sec.~3 of Ref.~\cite{mi001}. Within the conventional QFT framework based on the canonical formalism, each complete, momentum space, spin 0 Feynman propagator has the properties (a)-(c) but not (d) and (e), cf.~the footnote on p.~460 in Ref.~\cite{Weinberg}. 

\subsection{Particle content of an alternative scalar theory}
\label{partcont}

The QFT of a single, self-interacting real scalar field models scattering of scalar particles of the \it same mass. \rm Trying to find an alternative, finite theory to this QFT, we are interested in constructing an alternative, perturbative scattering matrix $\SMA$ that is a unitary model of quantum scattering of such particles. The masses of scalar particles whose quantum scattering is modelled by the $\SMA$-matrix are specified by the LSZ reduction formula. For their masses to be the same, the LSZ reduction formula requires that the analytic continuation $\GFAT(z)$, $z \in \CC$, of the alternative, momentum-space, two-point Green function $\GFAT(k^2)$, $k \in \RRR$, from real to complex values of $k^2$, is such that: (a)~$i\GFAT(z)$ is finite everywhere but at one point, say $z(\lambda)$; (b)~$i\GFAT(z)$ has a first-order pole at $z = z(\lambda)$; (c)~$z(\lambda)$ is finite and negative---the mass of scattered particles equals $\sqrt{ -z(\lambda)}$; and (d)~The corresponding residue is positive---otherwise $\SMA$ would not be unitary unless we introduced unphysical particles with negative metric, see Sects.~2.5 and 8 in Ref.~\cite{Hooft}; cf.~Eqs.~(10.3.17-18) in Ref.~\cite{Weinberg}. 

As $\GFAT(z)$ equals $\GFAT(k^2)$ for all real $z$, we have by (\ref{interLagr}) and (\ref{propagator}),
\begin{equation}
   \GFAT(z) = -i\fakt(z) / (z + m^2) \qquad\hbox{for all}\quad z \in \CC \qquad \hbox{if} \quad \lambda = 0  \,. 
   \label{limitselfenergy} 
\end{equation}
Thus, the alternative two-point Green function $\GFAT(z)$ has above properties (a)--(d) in the absence of interaction, i.e. for $\lambda = 0$, by (\ref{limitselfenergy}) and the properties of $\fakt(z)$. 

To consider whether $\GFAT(z)$ retains these properties in the presence of interactions, we write
\begin{equation}
   \GFAT(z) = -i\fakt(z) / [ z + m^2 - \SE(z, \lambda) ] \,, \qquad z \in \CC \,;
   \label{SMdef}
\end{equation}
by (\ref{limitselfenergy}) and (\ref{SMdef}),
\begin{equation}
   \SE(z, 0) = 0 \qquad \hbox{for all}\quad z \in \CC \,.
   \label{selfnointeraction}
\end{equation}
Let us show that $\GFAT(z)$ retains the above analytic properties (a)--(d) for sufficiently weak interactions if $\SE(z,\lambda)$ has the following two properties:

(A)~We can estimate that
\begin{equation}
   \sup_{z \in \CC} | \SE(z, \lambda) |  \to 0 \qquad \hbox{as} \quad \lambda \to 0 \,.
   \label{masscond}
\end{equation}

(B)~In a vicinity of $z = -m^2$ and $\lambda = 0$: (a)~$\SE(z, \lambda)$ is an analytic function of $z$ and $\lambda$, and (b)~$\Im \SE(z,\lambda) = 0 $ if $\Im z = 0$ and $\Im \lambda = 0$. 

Namely by (A), for each $\epsilon_1 > 0$ there is an $\epsilon_2 > 0$ such that $\GFAT(z)$ is finite if $| z + m^2 | \ge \epsilon_1$ and $| \lambda | \le \epsilon_2 $, by (\ref{SMdef}). Furthermore, each solution $z_i(\lambda)$ to the mass equation
\begin{equation}
   z = -m^2 + \SE (z,\lambda) \,,  \qquad z \in \CC \,,
   \label{massequat}
\end{equation}
either ceases to exist as $\lambda \to 0$, or
\begin{equation}
   z_i(\lambda) \to -m^2 \qquad \hbox{as} \qquad \lambda \to 0 \,,
   \label{zlamlim}
\end{equation}
because by (\ref{masscond}) and (\ref{massequat}) each solution $z_i(\lambda)$ is such that
\begin{equation}
   | m^2 + z_i(\lambda)|  \to 0 \qquad \hbox{as} \quad \lambda \to 0  \,.
\end{equation}
And as a consequence of (B), and of (\ref{selfnointeraction}) and its derivative, the implicit function theorem imples that: (i)~in a vicinity of $\lambda = 0$, there is only one solution, say $z(\lambda)$, to the mass equation (\ref{massequat}) that tends to $ -m^2$ as $\lambda \to 0$ (And by (\ref{zlamlim}), $z(\lambda)$ is also the only solution to the mass equation (\ref{massequat}) in a vicinity of $\lambda = 0$.); (ii)~$z(\lambda)$ is an analytic function of $\lambda$; (iii)~$z(\lambda)$ is real for real $\lambda$, since otherwise, $z^*(\lambda)$ would be an additional solution as $[\SE(z,\lambda)]^* = \SE(z^*,\lambda)$ if $\Im \lambda = 0$. So in a vicinity of $z = -m^2$ and $\lambda = 0$, $i\GFAT(z)$ is analytic with a first-order pole at $z = z(\lambda)$, and for real $\lambda$, the corresponding residue
\begin{equation}
   \fakt(z) \Bigl ( 1 - \partial \SE(z, \lambda) / \partial z \Bigr) ^{-1} \Big |_{ z = z(\lambda) }
   \label{residpos}
\end{equation}
is positive and tends to $1$ as $\lambda \to 0$, by (B), (\ref{selfnointeraction}), properties of $\fakt(z)$, and (\ref{zlamlim}).

On the analogy with QFT, one might take $\SE(z(\lambda), \lambda)$ as the alternative self-energy and $\sqrt{ - z(\lambda)}$ as the alternative renormalized mass of the scalar particle with mass $m = \sqrt{ - z(0)}$, by (\ref{massequat}) and (\ref{zlamlim}); cf.\ Refs.~\cite{Sterman}, \cite{Weinberg}, and \cite{Itzykson}.

\bf The perturbative expansion for $\SE(z, \lambda)$ in powers of $\lambda$ \rm is determined by the perturbative expansion of $\GFAT(z)$ in powers of $\lambda$ through (\ref{SMdef}). We may calculate it by a resumation of the perturbative expansion of $\GFAT(z)$ and so conclude that up to an arbitrary power of $\lambda$, $\SE(z, \lambda) $ is a power series in $\lambda$ whose coefficients are sums of truncated, one-particle-irreducible Feynman diagrams multiplied by $\fakt(z)$, cf.~e.g.~Eq.(10.3.14) in Ref.~\cite{Weinberg}. 

\subsection{Perturbative unitarity}
\label{perturunitar}

Following t'Hooft and Veltman (Sects.~4-7 in Ref.~\cite{Hooft}), one can show that the alternative perturbative $\SMA$-matrix is unitary up to any order in the coupling constant $\lambda$ provided: (i)~The Lagrangian specifying the alternative theory is Hermitian. (ii)~The corresponding alternative propagator $\prop(k^2)$ is such that (a)~it has a real spectral function and can be decomposed into positive and negative energy parts, and (b)~there are no ultraviolet divergencies. (iii)~The alternative, momentum-space two-point Green function $\GFAT(k^2)$ has properties (A) and (B) specified in the preceding Section and related to the particle content of the alternative theory. 

We have shown in Sec.~4 of Ref.~\cite{mi001} that there are alternative, Hermitian Lagrangians such that the corresponding alternative propagators $\prop(z)$ have the properties (\ref{propagator}) up to (\ref{faktest}), which imply the above properties in (ii). By maximum modulus theorem we can infer that the alternative propagator $\prop(z)$, given as an example in Sec.~4 of Ref.~\cite{mi001}, has also the property (\ref{ocenaodvoda}).

In the next section, we will show up to the second order in $\lambda$ that the properties (\ref{propagator})--(\ref{ocenaodvoda}) of the alternative propagator $\prop(k^2)$ suffice to make $\SE(z, \lambda)$ have properties (A) and (B) specified in Sec.~\ref{partcont}; and so they suffice to make the perturbative expansion of $\SMA$ unitary up to the same order.

\subsection{Properties of $\SE(z, \lambda)$ up to the second order in $\lambda$}
\label{pertprop}

By (\ref{propagator}), for the case considered the Wick rotation is possible, cf., e.g.~Sec.~9.2 in Ref.~\cite{Sterman}, and Ref.~\cite{Weinberg}. On assuming that 
\begin{eqnarray}
   Z &=& 1 - \lambda^2 Z_2 + \dots \,, \nonumber\\
   Z^2 \lambda_0 &=& \lambda + \lambda^2 L_2  + \dots \,, \label{sklopvrste} \\
   Zm_0^2 &=& m^2 - \lambda M_1 - \lambda^2 M_2 +\dots \,, \nonumber
\end{eqnarray}
we calculate that up to the second order in the coupling coefficient $\lambda$ the alternative self-energy
\begin{equation}
   \SE(q^2, \lambda) = \lambda A \fakt(q^2) + \lambda^2 [ B + I(q^2) - q^2 Z_2 ] \fakt(q^2) + \dots \,, 
   \label{selfenergy}
\end{equation}
where
\begin{eqnarray}
   A &=& M_1 - {1\over 2} \int {d^4 k \over (2\pi)^{4} } \,\prop(k^2) \,,  \label{inte2-0} \\
   B &=& M_2 - {1\over 2} A \int {d^4 k \over (2\pi)^{4} } \,\prop^2(k^2) 
                              - {1\over 2} L_2 \int {d^4 k \over (2\pi)^{4} } \,\prop(k^2) \,, \label{inte2-1}\\
   I(q^2) &=&  {1\over 6} \int {d^4 l \over (2\pi)^{4} } \int {d^4 k \over (2\pi)^{4} } \, \prop((k - q)^2) \prop((k- l)^2) \prop(l^2) \,, 
   \label{inte2}
\end{eqnarray}
$k$, $l$, $q \in \RRRR$, and $\epsilon=0$. That $I(q^2)$ and $\SE(q^2, \lambda)$ depend only on $q^2$ follows from the $\RRRR$-rotational invariance of alternative propagators in (\ref{inte2}), because for $q \in \RRRR$ all integrands in (\ref{inte2-0})--(\ref{inte2}) are absolutely integrable, by (\ref{faktest}). (The coefficients $M_1$, $Z_2$, $L_2$ and $M_2$ depend on the cut-off parameter $\Lambda$ in such a way that $\SE(q^2, \lambda)$ remains finite if we limit $\Lambda \to \infty$.)

Now we have to consider the analytic continuation $\SE(z, \lambda)$ of $\SE(q^2, \lambda)$ defined by (\ref{selfenergy})--(\ref{inte2}). To this end we use a complex variable $\zq \in \CC$ and replace $q \in \RRRR$ with 
\begin{equation}
   \qc \equiv (\zq, 0, 0, 0) 
   \label{qcdef}
\end{equation}
in rhs.(\ref{selfenergy}) and rhs.(\ref{inte2}). We then consider analytic properties of the function $\Iq(\zq)$, $\zq \in \CC$, which is defined by the rhs.(\ref{inte2}) with $q = \qc$, to verify that: (i)~$\Iq(\zq)$ depends in fact only on the complex variable 
\begin{equation}
   z = \zq^2 \,.
   \label{zzveza}
\end{equation}
(ii)~$\Iq(\zq)$ is such an extension of $I(q^2)$ that the corresponding $\SE(z,\lambda)$ satisfies requirements (A) and (B) of  Sec.~\ref{partcont} up to the second order in $\lambda$. We will introduce three transformations of Euclidean four-integrals in (\ref{inte2}) to construct forms suitable to this end.

\bf An estimate of $|\SE(z,\lambda) |$. \rm
Let us derive an estimate of $|\Iq(\zq)|$, valid for all $\zq \in \CC$. For a given $\zq = \zqr + i\zqi$, $\zqr, \zqi \in (-\infty, \infty)$, we introduce the subset $S_\alpha$ of Euclidean four-vectors $k = (k_0, \vec{k})$,
\begin{equation}
    S_\alpha \equiv \{  k_0 \in [ \zqr - \alpha, \zqr + \alpha] \,, \; 
                                     \vec{k}^2 \in [\zqi_2^2 - m^2 - \alpha^2, \zqi^2 - m^2 + \alpha^2] \} \subset \RRRR
   \label{setdefinition}
\end{equation}
with a positive parameter $\alpha < (- z_d - m^2)^{1/2}$; so for all $\qc$ we can estimate:
\begin{eqnarray}
   | \prop((k-\qc)^2) | &\le&  \alpha^{-2} a_0 \qquad \hbox{if} \qquad k \not\in S_\alpha \,, \nonumber \\
   \Re (k - \qc)^2  &>& z_d \qquad \hbox{if} \qquad k \in S_\alpha \,.    \label{racun20}
\end{eqnarray}
We integrate rhs.(\ref{inte2}) with $q = \qc$ by parts with respect to $k_0$ at each $k \in S_\alpha$. Using the characteristic function $\chi(k; \zq)$ that equals one if $k \in S_\alpha$ and zero otherwise, we construct the following result:
\begin{eqnarray}
   \Iq(\zq) &=& {1\over 6} \int {d^4 k \over (2\pi)^{4} } \int {d^4 l \over (2\pi)^{4} }   \Bigl \{ [ 1 - \chi(k; \zq)] \prop((k-\qc)^2) 
                \prop((k-l)^2)  \label{racun15} \\ 
            &&{}\kern- 50pt + \chi(k; \zq) \bigl [ (E_1 F_0 - E_0 F_1) ( \delta(k_0 - \zqr - \alpha) - \delta(k_0 - \zqr + \alpha) ) + E_0 F_2 ] \Bigr \}
                    \, \prop(l^2) \,,  \nonumber
\end{eqnarray}
where
\begin{eqnarray}
   E_n &=& {i \over 2 } (\vec{k}^2 + m^2)^{-1/2} \, {\partial^n\over\partial k_0^n} \bigl \{ (k_0 - q_-) \ln (k_0 - q_-) 
                             - (k_0 - q_+) \ln (k_0 - q_+)   \nonumber \\
            & & \qquad{}- [ (q_1 - q_-) \ln (q_1 - q_-) - (q_1 - q_+) \ln (q_1 - q_+) ] \bigr \} \,, \nonumber\\
   F_n &=& {\partial^n\over\partial k_0^n} \fakt((k - q_c)^2) \prop((k-l)^2) \,, \label{racun16} \\
   q_\pm &=& \zq \pm i (\vec{k}^2 + m^2)^{1/2} \,. \nonumber
\end{eqnarray}
Taking into account (\ref{racun20}), (\ref{ocenaodvoda}), and
\begin{eqnarray}
   \sup_{\zq \in \CC} \sup_{k \in S_\alpha} | E_0 |  &<& \infty \,, \\
   \sup_{\zq \in \CC} \sup_{k \in S_\alpha} | E_1 | \Big|_{ k_0= \zqr \pm \alpha } &<& \infty \,, 
\end{eqnarray}
we see that (i)~the integrand in (\ref{racun15}) is absolutely integrable (and so is the integrand in rhs.(\ref{inte2}) with $q = \qc$), (ii)
\begin{equation}
   \sup_{\zq \in \CC} |\Iq(\zq) | < \infty \,,
   \label{racun17}
\end{equation}
(iii)~we may change integration variables in rhs.(\ref{inte2}) with $q = \qc$, and (iv)
\begin{equation}
   \Iq(-\zq) = \Iq(\zq) \,,
   \label{iqsimetrija}
\end{equation}
by (\ref{propagator}) and (\ref{qcdef}), i.e., $\Iq(\zq)$ depends only on $z = \zq^2$.

So by (\ref{racun17}), (\ref{iqsimetrija}), (\ref{selfenergy})--(\ref{inte2}), and (\ref{propagator})--(\ref{faktest}), $\Iq(\zq)$ is such an extension of $I(q^2)$ that the self-energy $\SE(z,\lambda)$ is \it bounded up to the second order in the coupling constant $\lambda$ as required by condition (A) in Sec.~\ref{partcont}. \rm

\bf Analyticity of $\SE(z,\lambda)$. \rm
Applying Landau equations to (\ref{inte2}), we see that as long as $(\Im \zq)^2 < -z_d$, $\Iq(\zq)$ may have only four singularities: the ones at $\zq = \pm im$, and the ones at $\zq = \pm 3im$.

Let us show in four steps that $\Iq(\zq)$ is analytic in a vicinity of $\zq = \pm im$:

(A)~Introducing Feynman parameters in (absolutely integrable) rhs.(\ref{inte2}) with $q = \qc$ and changing integration variables, we get
\begin{equation}
   \Iq(\zq) = {1\over 6} \int {d^4 k \over (2\pi)^{4} } \int {d^4 l \over (2\pi)^{4} }  \int_0^1 d\alpha  \int_0^1 d\beta \,  
                   I_1(\alpha, \beta, k, l, \qc, m^2) \,,
   \label{racun1}
\end{equation}
where
\begin{eqnarray}
   I_1(\alpha, \beta, k, l, \qc, m^2 ) &=& 2\alpha [ D(\alpha, \beta, k^2, l^2, \qc^2, m^2) ]^{-3} \fakt(p_1^2) \fakt(p_2^2) \fakt(p_3^2 ) \,, \nonumber \\
   D(\alpha, \beta, k^2, l^2, z, m^2) &=& \alpha k^2 + b(\alpha, \beta) l^2 + c(\alpha, \beta) z + m^2 \,, \nonumber \\
   b(\alpha, \beta) &=& 1 - \alpha + \alpha \beta( 1 - \beta ) \in [0,1] \,,   \nonumber  \\
   c(\alpha, \beta) &=& \alpha \beta(1-\alpha)(1-\beta)/ b  \in [0 , 1/9] \,,    \label{racun4}\\
   p_1 &=& k + (1-\beta) l + [(1-\alpha)(1-\beta)/ b]  \qc \,,    \nonumber\\
   p_2 &=& l - [ \alpha\beta(1-\beta)/ b ] \qc \,,    \nonumber\\
   p_3 &=& k -\beta l - [ (1-\alpha)\beta / b ] \qc \,,    \nonumber
\end{eqnarray}
since $\qc^2 = z$. Note that
\begin{equation}
   (1-\alpha)(1-\beta) \,,\;  \alpha\beta(1-\beta) \,, \; (1-\alpha)\beta \; \in [0, b(\alpha,\beta)] \,.
   \label{racun8}
\end{equation}

(B)~By (\ref{racun4}), the expression $ D(\alpha, \beta, k^2, l^2, z, m^2) \ne 0 $ for each 
\begin{equation}
   z \in \SEP  \equiv \{ z:  | \Im z | \ge \epsilon' \} \cup \{ z: \Re z \ge -9 m^2 + \epsilon' \}
   \label{SEPdef}
\end{equation}
at any $\epsilon' > 0$. So the integrand $I_1(\alpha, \beta, k, l, \qc, m^2 )$ in rhs.(\ref{racun1}) is an analytic function of $\zq$ if $\zq^2 \in \SEP$ and $(\Im \zq)^2 < - z_d$ (since then $\Re p_i^2 > z_d$, $i = 1, 2, 3$, by (\ref{racun4}) and (\ref{racun8})).

(C)~We use (\ref{faktest}) and the estimate
\begin{equation}
   | (k - \qc)^2 + m^2 | \ge m^2 - (|z|  - \Re z )/2  \quad \hbox{for all}\quad k \in \RRRR,
\end{equation}
to infer that for $q = \qc$ the integrand in (\ref{inte2}) is absolutely integrable and
\begin{equation}
   | \Iq(\zq) | \le (16 \pi)^{-2} a_0^3  \big/ 6 [1 - (\Im z)^2 /2 m^2 (|z|  + \Re z ) ]
   \label{esti1}
\end{equation}
provided
\begin{equation}
   (\Im z)^2 < 2 m^2 (|z|  + \Re z ) \,.
   \label{noviem}
\end{equation}

(D)~In relations (\ref{inte2}), (\ref{racun1}), (\ref{esti1}), and (\ref{noviem}) we replace $m^2$ with
\begin{equation}
    m_1^2 = m^2 + | z | 
\end{equation}
to make (\ref{esti1}) and (\ref{noviem}) true for all $z \in \CC$. Then we take into account that for any $z \in \SEP$ we have
\begin{equation}
   r(z) \equiv \sup_{\alpha, \beta, k^2, l^2} | D(\alpha, \beta, k^2, l^2, z, m_1^2)/ D(\alpha, \beta, k^2, l^2, z, m^2) |^3 < \infty 
   \label{RMdef}
\end{equation}
to infer that for all $z \in \SEP$: (a)~the integrand $I_1(\alpha, \beta, k, l, \qc, m^2 )$ in (\ref{racun1}) is absolutely integrable, and (b)~we can estimate that
\begin{equation}
   | \Iq(\zq) | \le (16 \pi)^{-2} a_0^3  r(z) (m^2 + z) / 6 [ m^2 + (| z |  + \Re z )/2 ] \,.
\end{equation}
By (a), (\ref{racun8}), Cauchy's integral representation of an analytic function, and Fubini's theorem, we can infer from (\ref{racun1}) that $\Iq(\zq)$ is an analytic function of $\zq$ provided $\zq^2 \in \SEP$ and $(\Im \zq)^2 < - z_d$. So
\begin{equation}
   \Im \Iq(\zq) = 0 \qquad \hbox{for all} \quad \zq^2 > -9m^2 \,,
   \label{realnost}
\end{equation}
since $\Iq(\zq)$ is analytic for all $\zq^2 > -9m^2$ and since $\Iq(\zq) = I(q^2)$ and $\Im I(q^2) = 0$ for all $q^2 \ge 0$, by (\ref{inte2}).

As a consequence, $\Iq(\zq)$ is such an analytic extension of $I(q^2)$ that \it the self-energy $\SE(z, \lambda)$ satisfies condition (B) of  Sec.~\ref{partcont} up to the second order in $\lambda$, \rm by (\ref{selfenergy})--(\ref{inte2}), (\ref{propagator})--(\ref{faktest}), (\ref{iqsimetrija}), and (\ref{racun1})--(\ref{realnost}).

\section{Comments}
\label{komentar}

We have shown in Sec.~\ref{pertprop} that the analytic properties (\ref{propagator})--(\ref{ocenaodvoda}) of the alternative propagator $\prop(k^2)$ endow the alternative self-energy $\SE(z,\lambda)$ with the properties (A) and (B) in Sec.~\ref{partcont} at least up to the second order in the coupling constant $\lambda$. So the alternative scattering matrix $\SMA$ is unitary at least up to the same order.

The question remains whether the analytic properties (\ref{propagator})--(\ref{ocenaodvoda}) of $\prop(k^2)$ suffice for proving the unitarity of $\SMA$ to all orders in $\lambda$.

That they suffice to all orders in $\lambda$ to make $\SE(z,\lambda)$ such a bounded function as required by condition (A) in Sec.~\ref{partcont} can be seen as follows. First we modify each diagram so that: (i)~it has the least possible number of internal lines that carry the external momentum $q$, and (ii)~each such line carries only one internal momentum. Then we estimate each so modified diagram on the analogy to the method (\ref{setdefinition})--(\ref{iqsimetrija}) we used to estimate the setting sun diagram (\ref{inte2}).

They likely suffice also to make $\SE(z,\lambda)$ to all orders in $\lambda$ such an analytic function as required by condition (B) in Sec.~\ref{partcont}, but we lack a formal proof.

\section*{Acknowledgement}

We would like to thank Matja\v z Polj\v sak for many useful discussions.

\end{document}